\documentclass[twocolumn,showpacs,preprintnumbers,amsmath,amssymb]{revtex4}

\usepackage{amsmath}
\usepackage{graphicx}

\begin{document}

\title{Rotational optomechanical coupling of a spinning dielectric sphere}


\author{H. K. Cheung and C. K. Law}
\affiliation{Department of Physics and Institute of Theoretical
Physics, The Chinese University of Hong Kong, Shatin, Hong Kong SAR,
China}

\date{\today}
\begin{abstract}

We formulate a non-relativistic Hamiltonian in order to describe
how the rotational degrees of freedom of a dielectric sphere and
quantized light fields are coupled. Such an interaction is shown
to take a form of angular momentum coupling governed by the field
angular momentum inside the dielectric. As a specific example, we
show that the coupling due to a single whispering gallery mode can
lead to precession dynamics and frequency shifts of light.

\end{abstract}

\pacs{42.50.Tx, 42.50.Wk, 42.50.Pq, 45.20.dc}

\maketitle

Cavity quantum optomechanics has been an active research area
investigating quantum phenomena and applications through the
interaction between mechanical and optical degrees of freedom
\cite{Kippenberg rev 1, Kippenberg rev 2, Favero rev, Grivin rev,
Kiesel rev}. In particular, since the mechanical systems such as a
dielectric membrane or sphere have masses much greater than that
of an atom, the study of quantum optomechanics may test the
foundation of quantum theory in macroscopic systems
\cite{Bouwmeester2003}. Typically, the systems considered in
cavity quantum optomechanics are deformable cavities. These
cavities are subject to radiation pressure pushing their cavity
`walls' apart, which in turn changes the field dynamics. For such
systems, the deformation of the cavity is fundamental to the
strong coupling between optics and mechanics.

In this paper, we discuss another type of optomechanical coupling
that results from the {\em rotation} of the optical cavity, in
which case the coupling remains even in the absence of cavity
deformation. Physically, an optical field can affect the
rotational motion via the electromagnetic torque exerting to a
dielectric object \cite{Marton,Lee,Loudon}, and this has been
studied in a sequence of experiments
\cite{Beth,Padgett,Padgett2,Dunlop,Paterson}. The mechanical
rotation in turn affects light inside the dielectric, not only
because the dielectric changes its orientation, but also due to
the {\em motional-induced} polarization and magnetization
\cite{Landau}. Such a rotational optomechanical coupling could
lead to a non-trivial coupled dynamics, and it is then a natural
question on how the problem can be formulated self-consistently.
In particular, a Hamiltonian formalism of the system would allow a
generalization to a fully quantized theory, in which both the
optical and mechanical degrees of freedom are quantized.

We remark that there are recent studies beginning to explore
quantum effects in optically-trapped dielectric sphere
\cite{kimble nanosphere,raizen,cirac}. Since the orientation of
the levitated dielectric particle is not fixed, rotational
dynamics could be possible. However, it remains unclear about the
strength of rotational coupling and what fundamental effects can
be produced when light is acting on a rotating sphere and vice
versa.

The goal of this paper is to formulate a Hamiltonian that can
address the rotational dynamics of the coupled dielectric-field
system. Specifically, we consider a rigid dielectric sphere with
radius $R$ and moment of inertia $I$, placed in free space. The
dielectric constant of the sphere is given by
\begin{equation}
\epsilon(\bf r)
=
 \left\{ \begin{array}{ll}
            n^2,& |{\bf r} - {\bf r}_0| \leq R \\
            1,& \mbox{otherwise}.
          \end{array}
 \right.
\end{equation}
We have used the convention $\epsilon_0 = \mu_0 = 1$ (i.e.,
$c=1$), and assumed non-magnetic dielectric $\mu = \mu_0$. We have
also assumed a non-dispersive and non-absorptive dielectric.

The sphere is free to rotate about any axis, but its
center-of-mass (CM) is {\em fixed} at ${\bf r}_0$. In practice the
sphere may be confined by an external potential so that the CM of
the sphere moves about an equilibrium position. We assume such
motion to be negligibly slow and of negligible amplitude, then
since for a spherical object, the CM motion does not directly
couple with the rotational degrees of freedom \cite{cirac}, our
approach here would be a good approximation. The system is
specified by the Lagrangian
\begin{equation}
L = \frac{1}{2} I {\boldsymbol \omega}^2 + \int d^3{\bf r} {\cal L}(\bf r)
\end{equation}
where ${\boldsymbol \omega} = (\dot{\gamma} \sin\beta \cos\alpha -
\dot{\beta}\sin\alpha) \hat{\bf x} +(\dot{\gamma} \sin\beta
\sin\alpha + \dot{\beta}\cos\alpha) \hat{\bf y} + (\dot{\alpha} +
\dot{\gamma} \cos\beta) \hat{\bf z}$ is the angular velocity of
the sphere, $\alpha, \beta, \gamma$ are the 3 Euler angles
specifying the orientation of the sphere (we follow the convention
in Ref.~\cite{euler ang}). ${\cal L}$ is the Lagrangian density of
the field after eliminating the electronic degrees of freedom of
the dielectric. To find ${\cal L}$, we go to an {\em inertial}
frame $S'({\bf r})$ in which the dielectric element at $\bf r$ is
instantaneously at rest. Assuming the acceleration of the
dielectric does not change its macroscopic properties, the field
Lagrangian density at $\bf r$ in $S'({\bf r})$ is given by the
familiar form: ${\cal L}' = \frac{1}{2}\left( \epsilon {\bf E}'^2
- {\bf B}'^2 \right)$, where ${\bf E}'$ and ${\bf B}'$ are the
electric and magnetic fields in $S'({\bf r})$, respectively. As
the Lagrangian density is Lorentz-invariant, ${\cal L}$ can be
readily obtained from the Lorentz transformation of the fields
from $S'({\bf r})$ to the laboratory frame $S$. We confine
ourselves to a non-relativistic motion of the sphere, so that the
velocity ${\bf v}({\bf r}) ={\boldsymbol \omega}\times({\bf r} -
{\bf r}_0)$ of the dielectric element at any point $\bf r$
satisfies $|{\bf v}({\bf r})|\ll c$. Linearizing ${\cal L}$ up to
first order on $\boldsymbol \omega$, the Lagrangian reads
\begin{eqnarray}
L =\frac{1}{2} I {\boldsymbol \omega}^2
    + \frac{1}{2} \int d^3{\bf r} \left(\epsilon {\bf E}^{2} - {\bf B}^{2}\right)
    - {\boldsymbol \omega}\cdot{\bf \Gamma},
\label{eq:L in terms of omega}
\end{eqnarray}
where ${\bf \Gamma} = \int d^3{\bf r} (\epsilon-1)({\bf r} - {\bf
r}_0) \times\left({\bf E}\times{\bf B}\right)$, which takes a form
similar to the field angular momentum stored in the sphere. The
Lagrangian (\ref{eq:L in terms of omega}) is a generalization to
that of Barton {\em et al.}~\cite{Barton} and
Salamone~\cite{salamone} (in the case $\mu=1$), which consider a
one-dimensional configuration and focus on CM motion of a
dielectric slab. Here we will take ${\boldsymbol \omega}$ as a
degree of freedom which interacts with the field through the
$-{\boldsymbol \omega}\cdot{\bf \Gamma}$ term.

The electromagnetic field is specified by the scalar potential
$V({\bf r},t)$ and vector potential ${\bf A}({\bf r},t)$ under the
{\em generalized radiation gauge} $\nabla\cdot\left[\epsilon({\bf
r}) {\bf A}\right] = 0$ in the presence of dielectric \cite{glauber,
knight, knoll}, with ${\bf E} = -\left(\partial_t {\bf A}\right) -
\nabla V$ and ${\bf B} = \nabla\times {\bf A}$. For the completeness
of our theory, let us first discuss the Euler-Lagrange equations of
the system before going to the Hamiltonian.

The first Euler-Lagrange equation for the field is a restatement
of the Maxwell equation $\nabla\cdot{\bf D}=0$, which reads: $
\nabla \cdot \left(\epsilon \nabla V\right) = \nabla \cdot \left[
\left(\epsilon-1\right){\bf v} \times\left(\nabla\times{\bf
A}\right) \right]$. This is understood from the fact that the
polarization of a moving dielectric element is ${\bf P} =
(\epsilon-1)\left({\bf E} + {\bf v}\times{\bf B}\right)$ to first
order of $v\equiv|{\bf v}|$. Under the generalized radiation
gauge, $V$ is not a degree of freedom and it is determined by the
{\em instantaneous} values of ${\bf A}({\bf r},t)$ and ${\bf
v}({\bf r},t)$. We see that $V$ is linear in $\boldsymbol \omega$,
and vanishes when the dielectric is at rest. The second
Euler-Lagrange equation of the field is the wave equation: $
\partial_t \left(\epsilon \partial_t{\bf A}\right)
+ \nabla\times \left(\nabla \times {\bf A}\right) = {\bf j}, $
where ${\bf j}$ is a motion-induced (i.e. ${\cal O}(v)$) source
current density,
\begin{eqnarray}
{\bf j} = \partial_t \left[ -\epsilon \nabla V + \left(\epsilon -1\right)
{\bf v} \times{\bf B} \right] +\nabla \times {\bf M}.
\end{eqnarray}
The wave equation is consistent with the Maxwell equation
$\nabla\times{\bf B} =\partial_t{\bf D} + \nabla\times{\bf M}$ in
which ${\bf M} = -{\bf v} \times {\bf P}$ is the magnetization of
a moving dielectric with the polarization $\bf P$. From these two Euler Lagrange equations, we see that the
motion-induced source terms are fundamental to the sphere-field
coupling, without which the rotation of the dielectric sphere
cannot affect the time evolution of the field.

The mechanical equation of motion follows from the Euler-Lagrange
equations of the Euler angles $(\partial L/\partial \zeta) =
(d/dt) (\partial L/\partial \dot{\zeta})$, $\zeta =
\alpha,\beta,\gamma$. In terms of $\boldsymbol \omega$, we have
\begin{equation}
I \frac{d \boldsymbol \omega}{dt} = - {\boldsymbol \omega}\times
{\bf \Gamma}+  \frac{d \bf \Gamma}{dt} \label{eq:EOM omega}.
\end{equation}
The two terms on the RHS of Eq.~(\ref{eq:EOM omega}) characterize
two different types of dynamics of $\boldsymbol \omega$. The first
term describes a precession about the $\bf \Gamma$ axis with a
frequency $|{\bf\Gamma}|/I$, which keeps the magnitude of
$\boldsymbol \omega$ unchanged \cite{precession}. On the other
hand, the second term may change the magnitude of $\boldsymbol
\omega$ along the $\bf \Gamma$ axis. We remark that
Eq.~(\ref{eq:EOM omega}) is consistent with the conservation of
total angular momentum $\frac{d}{dt}(I{\boldsymbol \omega} + {\bf
J}_F)=0$ which follows from the rotational invariance of the
Lagrangian (\ref{eq:L in terms of omega}). To zeroth order in
$\boldsymbol \omega$, the total field angular momentum reads ${\bf
J}_F = \int d^3{\bf r} ({\bf r} - {\bf r}_0) \times\left({\bf
E}\times{\bf B}\right)$ \cite{Field J remark}.

We now turn to the Hamiltonian defined from $L$ by
\begin{equation}
H \equiv \sum_{\zeta=\alpha,\beta,\gamma} \dot{\zeta} p_\zeta
+ \int {\bf \Pi}\cdot\left(\partial_t{\bf A}\right) d^3 {\bf r} - L,
\label{eq:H_def}
\end{equation}
where ${\bf \Pi}({\bf r},t) \equiv [\partial {\cal L}/ \partial
(\partial_t{\bf A})]$ is the field canonical momentum density, and
$p_\zeta \equiv (\partial L/\partial \dot{\zeta})$ are the
canonical momenta conjugate to the Euler angles. We introduce a
{\em canonical angular momentum} $\bf J$ in terms of $p_\zeta$
\cite{remark J}:
\begin{eqnarray}
J_x
    &=& -\cot\beta \cos\alpha p_{\alpha}
    - \sin\alpha p_{\beta}
    + \csc\beta \cos\alpha p_{\gamma}, \nonumber\\
J_y
    &=& -\cot\beta \sin\alpha p_{\alpha}
    + \cos\alpha p_{\beta}
    + \csc\beta \sin\alpha p_{\gamma}, \nonumber\\
J_z
    &=& p_{\alpha}.
\label{eq: J def euler}
\end{eqnarray}
Explicitly, ${\bf \Pi}({\bf r},t)$ and $\bf J$ are given by
\begin{eqnarray}
{\bf \Pi} &=& -\epsilon {\bf E} - (\epsilon-1) \left({\bf
v}\times{\bf B}\right) = -{\bf D},
\label{eq:Pi def}\\
{\bf J} &=&  I {\boldsymbol \omega} - {\bf \Gamma}.
\label{eq:J_def omega}
\end{eqnarray}
Note that ${\bf \Pi}$ is transverse as $\nabla\cdot{\bf D}=0$, and
$\bf J$ differs from the kinetic angular momentum $I {\boldsymbol
\omega}$ for non-zero fields. The explicit expression of the
Hamiltonian (\ref{eq:H_def}) reads
\begin{eqnarray}
H
=  \frac{\left({\bf J} + {\bf \Gamma}'\right)^2}{2I}
+ \frac{1}{2} \int d^3{\bf r}\left(\frac{{\bf \Pi}^2}{\epsilon} + {\bf B}^2\right),
\label{eq:H w/o CM}
\end{eqnarray}
with ${\bf \Gamma}'$ given by
\begin{eqnarray}
{\bf \Gamma}' &=& - \int d^3{\bf r}\left(\frac{\epsilon-1}{\epsilon}\right)
            \left[({\bf r}-{\bf r}_0)\times\left({\bf \Pi}\times{\bf B}\right)\right].
\label{eq: Gamma' def}
\end{eqnarray}
This Hamiltonian takes a form similar to the minimal-coupling
Hamiltonian in electrodynamics, with ${\bf \Gamma}'$ somehow
playing the role of vector potential in the kinetic energy term.
We note that in writing Eq.~(\ref{eq:H w/o CM}), we have neglected
field-dependent terms that are quadratic in ${\bf v}$. These terms
resemble the kinetic energy of the sphere, and contribute to a
correction of the moment of inertia $I$ due to the field. Such a
correction is typically very small compared with $I$ for fields
well below the dielectric breakdown of the sphere.

With the classical Hamiltonian (\ref{eq:H w/o CM}), the canonical
quantization of the system is readily achieved by promoting the
dynamical variables $\zeta$, $p_\zeta$ ($\zeta =
\alpha,\beta,\gamma$), ${\bf A}({\bf r})$ and ${\bf \Pi}({\bf r})$
into operators by postulating the commutation relations:
\begin{eqnarray}
&& [\zeta, p_\eta] = i\hbar \delta_{\zeta \eta} \\
&& [A_i({\bf r}), \Pi_j({\bf r}')] = i\hbar
\delta^{\epsilon}_{ij}({\bf r},{\bf r}') \end{eqnarray} where
$\delta^{\epsilon}_{ij}({\bf r},{\bf r}')$ is a {\em generalized
transverse $\delta$-function} in the presence of dielectric
\cite{glauber}. From Eq.~(\ref{eq: J def euler}), the commutation
relations of the Euler angles $\alpha,\beta,\gamma$ imply that
$\bf J$ forms a quantum rigid rotor \cite{euler ang}, which
includes the angular momentum commutation relation $[J_i, J_j] =
i\hbar \epsilon_{ijk} J_k$. The quantum Hamiltonian takes the same
expression as Eq.~(\ref{eq:H w/o CM}), but with ${\bf \Gamma}'$
defined in Eq.~(\ref{eq: Gamma' def}) symmetrized, i.e. with the
bracketed term in the integrand replaced by $\left[({\bf r}-{\bf
r}_0)\times\left({\bf \Pi}\times{\bf B} - {\bf B}\times{\bf
\Pi}\right) \right]/2$.

In order to discuss field excitations in Fock space, we project
the field operators onto a complete set of mode functions, namely
the TE and TM mode functions (using spherical coordinates with
origin at ${\bf r}_0$) $ {\bf \Psi}_{lm}(k, {\bf r}) =
u^{(E)}_{l}(k,r) {\bf X}_{lm}(\theta,\phi) $ and $ {\bf
\Phi}_{lm}(k, {\bf r}) =(i/k)\nabla \times [u^{(M)}_l(k,r) {\bf
X}_{lm}(\theta,\phi)] $ respectively, with ${\bf
X}_{lm}(\theta,\phi)$ being the vector spherical harmonics
\cite{Jackson}, and $u^{(E)}_{l}(k,r)$ and $u^{(M)}_{l}(k,r)$ are
radial functions subject to appropriate boundary conditions across
$r=R$: $\nabla\cdot\left[\epsilon(r){\bf \Psi}_{lm}(k, {\bf
r})\right] = \nabla\cdot\left[\epsilon(r){\bf \Phi}_{lm}(k, {\bf
r})\right] =0$ \cite{Jaewoo}. Substituting the normal-mode
expansion into the Hamiltonian (we take $\hbar =1$ from here on),
\begin{eqnarray}
H &=& \frac{\left({\bf J}+{\bf \Gamma}'\right)^2}{2I} + H_F
\label{eq: total H w/ mode}
\end{eqnarray}
where $H_F = \int dk \sum_{l,m} \omega_{k}
[a^{\dag}_{lm}(k)a_{lm}(k)+ b^{\dag}_{lm}(k)b_{lm}(k)]$ is the
field Hamiltonian with constant terms removed, $a_{lm}(k)$ and
$b_{lm}(k)$ are the annihilation operators for TE and TM mode
photons with quantum numbers $(k,l,m)$ and a frequency $\omega_k =
ck$, respectively.

Equations (\ref{eq:H w/o CM}) and (\ref{eq: total H w/ mode}) are
main results of this paper. It is important to note that the form of
${\bf \Gamma}'$ [Eq.~(\ref{eq: Gamma' def})] is very similar to the
field angular momentum stored in the dielectric, apart from some
proportionality constant. Therefore, approximately speaking, the
first term of the Hamiltonian (\ref{eq:H w/o CM}) and (\ref{eq:
total H w/ mode}) represents an angular momentum coupling, i.e., the
interaction corresponds to an exchange of angular momenta between
the field and the sphere. In particular, if the field is localized
inside the sphere, we expect that ${\bf \Gamma}'$ should become a
good approximation to the {\em total} field angular momentum, up to
a multiplicative constant.

As an illustrative example of the rotational optomechanical
coupling, we apply the Hamiltonian (\ref{eq: total H w/ mode}) to
a configuration in which photons occupy a whispering-gallery mode
(WGM). In this case photons can be confined inside the dielectric
cavity with a long life time due to multiple total internal
reflections. For simplicity, we consider that the field excitation
is dominantly contributed by TE mode photons with frequencies
$\omega_k\approx \omega_0$,  where $\omega_0=ck_0$ is a resonant
frequency of a TE WGM. Assuming that such a WGM has a narrow line
width $\kappa_c$ and $\omega_0$ is well separated from all other
TM mode frequencies, it is sufficient to include TE modes only in
the Hamiltonian. Furthermore, since the optical quality factor of
the spherical cavity is typically very high, i.e. $Q =
k_0/\kappa_c\gg1$, it is instructive to consider the dynamics of
the system within a time scale short compared with
$\kappa_c^{-1}$. In this regime, the leakage of WGM photons is
negligible. Then the field Hamiltonian associated with TE WGMs
with orbital quantum number $l$ is given by
$H_F\approx\sum_{m=-l}^{l}\omega_{0} c^{\dag}_{m}c_{m}$, where
$c_{m}$ is the cavity mode operator~\cite{remark_mode}
\begin{eqnarray}
c_{m} = \sqrt{\frac{\kappa_c}{\pi}} \int dk \frac{a_{lm}(k)}{k-k_0 +
i\kappa_c}. \label{eq: c_m def}
\end{eqnarray}
Here the index $l$ for $c_m$ is suppressed for compactness.

With the help of cavity mode $c_{m}$ operators, the $\bf {\Gamma}'$
operator contributed by the TE WGM is approximately given by,
\begin{eqnarray}
{\bf \Gamma}' \approx \Lambda\sum_{mm'}\left(\int d\Omega
Y^{*}_{lm'}{\bf L}Y_{lm}\right)  c^{\dag}_{m'}c_{m} \equiv \Lambda
{\bf S}, \label{eq:Gamma'}
\end{eqnarray}
where ${\bf L} = -i({\bf r}-{\bf r}_0)\times\nabla$,
$Y_{lm}(\theta,\phi)$ are spherical harmonics and
\begin{eqnarray} \Lambda = \pi \kappa_c \int_{0}^{R} dr
(\epsilon-1) r^2 |u^{(E)}_{l}(k_0,r)|^2
\end{eqnarray}
is a dimensionless parameter determined by the mode amplitude
inside the sphere. Numerical calculations with the parameters $R=
10$\mbox{ $\mu$m}, $n^2 = 2.31$, $l = 120$, $k_0 = 2\pi/(743.25
\mbox{ nm})$ leads to $\Lambda=1.12$.

In writing Eq.~(\ref{eq:Gamma'}) we have employed the
rotating-wave approximation (RWA), so that fast oscillating terms
such as $c^{\dag}_{m}c_{m'}^{\dag}$ are dropped. However, these
terms are responsible for photon generation in dynamical Casimir
effect \cite{Dodonov}, and they should be retained if such a
quantum effect become significant, for example, when ${\boldsymbol
\omega}(t)$ is rapidly changing with time. We also remark that the
angular integral in Eq.~(\ref{eq:Gamma'}) gives the selection
rules for the rotational coupling. Together with $[c_m,
c_{m'}^{\dag}]=\delta_{mm'}$, we see that $\bf S$ defined in
Eq.~(\ref{eq:Gamma'})  satisfies the angular momentum commutation
relations $[S_i, S_j] = i\epsilon_{ijk} S_k$.

By combining Eq.~(\ref{eq: total H w/ mode}), the interaction with
the TE WGM leads to a Hamiltonian (in a rotating frame where $H_F$
is eliminated),
\begin{eqnarray}
H_r = \Lambda \frac{\left({\bf J}+{\bf S}\right)^2}{2I}
    + \left(1-\Lambda\right) \frac{{\bf J}^2}{2I}
    + \Lambda\left(\Lambda-1\right) \frac{{\bf S}^2}{2I}.
\label{eq:H rotating frame}
\end{eqnarray}
which describes a coupling between the angular momenta $\bf J$ and
$\bf S$, with coupling strengths characterized by $\Lambda$. In
addition, the Heisenberg equations of motion for $\bf S$ and
${\boldsymbol \omega}$ (noting that ${\bf J} = I{\boldsymbol
\omega} -\Lambda {\bf S}$) become
\begin{eqnarray}
\dot{\bf S} = \Lambda {\boldsymbol \omega}\times{\bf S}
\hspace{5mm}\mbox{and}\hspace{5mm}
I\dot{\boldsymbol \omega} = \Lambda(\Lambda-1)
\left({\boldsymbol \omega} \times {\bf S}\right)
\label{eq: WGM EOM}
\end{eqnarray}
respectively, indicating that the optical ($\bf S$) and mechanical
angular momentum ($I{\boldsymbol \omega}$) precess about each
other in the rotating frame. Moreover, for the coherent time scale
concerned here (i.e., with negligible cavity field decay), the
gallery mode photons cannot change the magnitude of $\boldsymbol
\omega$, and the sphere can only precess about the instantaneous
$\bf S$-axis.

An order-of-magnitude estimate of the mechanical precession rate
can be made by $\Lambda(\Lambda-1) \langle {\bf S} \rangle /I
\approx (n^2-1)N\hbar l/\rho R^5$, where $N$ is the number of
cavity photons, and $\rho$ is the mass density of the sphere.
Under the same numerical parameters above, and assuming $N\approx
10^{5}$ before dielectric breakdown, the precession rate would be
on the order of $10^{-5}\mbox{ Hz}$. If the sphere spins {\em
coherently} at a {\em macroscopic} rate and the WGM field inside
the cavity is sufficiently weak such that $I|\langle{\boldsymbol
\omega}\rangle| \gg |1-\Lambda||\langle{\bf S}\rangle|$, then
${\boldsymbol \omega}\approx \langle{\boldsymbol \omega}\rangle$
behaves approximately as a constant (classical) vector. In this
case the field dynamics described by $\bf S$ is governed by an
effective Hamiltonian $H_{\rm eff} = \Lambda \langle{\boldsymbol
\omega}\rangle\cdot {\bf S}$, which has the same form as that of a
magnetic moment in an external magnetic field. Therefore photons
initially occupying the cavity mode $m$ experiences a {\em
Zeeman-type} frequency shift of $m\Lambda\langle\omega_z\rangle$.
We note that the frequency shift should be resolved from the
cavity linewidth (i.e.
$\Lambda\langle\omega_z\rangle\gtrsim\kappa_c$) in order to be
observable. With an optical field with quality factor of $Q\sim
10^{10}$, such a condition would require the dielectric sphere to
spin at a rate of $1\mbox{ kHz}$.

To conclude, we have established a non-relativistic Lagrangian and
a Hamiltonian for a three-dimensional sphere-field system in the
$v \ll c$ regime. By including the motion-induced polarization and
magnetization possessed by the dielectric sphere, we have
self-consistently determined how a spinning dielectric sphere and
quantized light fields are coupled. The sphere-field interaction
is described by the Hamiltonian (\ref{eq:H w/o CM}) and (\ref{eq:
total H w/ mode}) as a coupling of the canonical angular momentum
$\bf J$ to the quantity ${\bf \Gamma}'$, which is proportional to
the field angular momentum stored in the dielectric sphere. We
further illustrated the rotational coupling using a WGM
description of the Hamiltonian, and identified the constant
$\Lambda$ that determines the sphere-field coupling strength.
Within the coherent time scale of the WGM photons, we have shown
that the optical and mechanical angular momentum precesses about
each other, and the degenerate WGM multiplets would experience a
Zeeman-type splitting under a strong mechanical rotation. Although
such effects are weak in general, they reveal fundamental features
arising from the rotational degrees of freedom of the fields and
the sphere. Our work here should provide a framework to further explore quantum phenomena and applications of such rotational
optomechanical coupling.

\emph{Acknowledgments}--This work is partially supported by a
grant from the Research Grants Council of Hong Kong, Special
Administrative Region of China (Project No.~CUHK401810).


\begin{thebibliography}{99}
\bibitem{remark J}
$\bf J$ can be equivalently defined as ${\bf J} = (\partial
L/\partial {\boldsymbol \omega})$. Furthermore, it can be shown
that ${\bf J}\cdot {\boldsymbol \omega} = \dot{\alpha} p_\alpha +
\dot{\beta} p_\beta + \dot{\gamma} p_\gamma$, so that $H$ in
Eq.~(\ref{eq:H_def}) can be formulated using $\bf J$ directly.
\bibitem{remark_mode}
The definition of $c_{m}$ is based on the transmission coefficient
of TE mode function $T^{(E)}_{l}(k)$ (see \cite{Jaewoo}), whose
analytic continuation possesses well-separated simple poles, so
that $T^{(E)}_{l}(k) \approx |T^{(E)}_{l}(k_0)|\kappa_c / (k- k_0
+ i\kappa_c)$ near a simple pole at $k=k_0 - i\kappa_c$. The
gallery-mode operator $c_{m}$ can then be constructed as
superpositions of $a_{lm}(k)$ using the Lorentzian form of
$T^{(E)}_{l}(k)$ as given in Eq.~(\ref{eq: c_m def}).
\end{thebibliography}
\end{document}